\documentstyle[12pt,aaspp4]{article}
\begin{document}

\title{HISTORY OF THE LOCAL GROUP}

\author{SIDNEY van den BERGH}

\affil{Dominion Astrophysical Observatory\\
 Herzberg Institute of Astrophysics\\ 
National Research Council of Canada\\
5071 West Saanich Road, Victoria, BC, Canada  V9E 2E7\\
email: sidney.vandenbergh@nrc-cnrc.gc.ca}

\begin{abstract}

It is suggested that M31 was created by the early merger, and 
subsequent violent relaxation, of two or more massive metal-rich
ancestral galaxies within the core of the Andromeda subgroup of the
Local Group. On the other hand the evolution of the main body of 
the Galaxy appears to have been dominated by the collapse of a single 
ancestral object, that subsequently evolved by capturing a halo of 
small metal-poor companions. It remains a mystery why the  globular 
cluster systems surrounding galaxies like M33 and the LMC exhibit 
such striking differences in evolutionary history. It is argued that 
the first generation of globular clusters might have been formed 
nearly simultaneously in all environments by the strong pressure 
increase that accompanied cosmic reionization. On the other hand 
subsequent generations of globulars may have formed during 
starbursts that were triggered by collisions and mergers of gas rich galaxies. 
\end{abstract}

 \textit{The fact that the [G]alactic system is a member of a group is a very 
fortunate accident.} Hubble (1936, p.125)

\section{INTRODUCTION}
   According to Greek mythology the goddess of wisdom, Pallas Athene, 
emerged clad in full armor after Hephaestus split open Zeus's
head. In much the same way the Local Group sprang forth suddenly,
and almost complete, in Chapter VI of The Realm of the Nebulae
(Hubble 1936, pp. 124-151). Hubble describes the Local 
Group
as ``a typical small group of nebulae which is isolated in the
general field''. He assigned (in order of decreasing luminosity) M31, 
the Galaxy, M33, the Large Magellanic Cloud, the Small Magellanic 
Cloud, M32, NGC205, NGC 6822, NGC 185, IC 1613 and NGC 147 to the
Local Group, and regarded IC 10 as a possible member. In the 
2/3 century since Hubble's work, the number of known Local Group
members has increased from 12 to 36 (see Table 1) by the addition
of almost two dozen low-luminosity galaxies. Recent detailed 
discussions of individual Local Group galaxies are given in Mateo (1998), Grebel (2000) and van den Bergh (2000a). Hubble (1936, p. 128)
pointed out that investigations of the Local Group were important for 
two reasons: [1] ``[T]he members have been studied individually, as 
the nearest and most accessible examples of their particular types, 
in order to determine the[ir] internal structures and stellar 
contents''. In the second place [2], ``the [G]roup may be examined as 
a sample collection of nebulae, from which criteria can be derived 
for further exploration''. 

   Small galaxy groups, like the Local Group, are quite common. From
inspection of the prints of the {\it Palomar Sky Survey} van den 
Bergh (2002a) has estimated that 16\% of nearby galaxies are 
located in
such small groups. Hubble (1936, p. 128) 
emphasized that ``The groups
[such as the Local Group] are aggregations drawn from the general
field, and are not additional colonies superposed on the field''.
From its observed radial velocity dispersion of 61 $\pm$ 8 km $s^{-1}$ 
the
Local Group is found to have a virial mass of (2.3 $\pm$ 0.6) x 
$10^{12}$ 
M$\odot$ (Courteau \& van den Bergh 1999). The 
zero-velocity surface of 
the Local Group has a radius of 1.18 $\pm$ 0.15 Mpc, but $\sim$80\% of 
the
Local Group members are actually situated within 0.4 Mpc of the
barycenter of the Local Group, which is located between M31 and 
the Galaxy. From its virial mass, and the integrated luminosity 
of Group members of 4.2 x 10 $^{10}$ L$\odot$, the dynamical 
mass-to-light 
ratio of the Local Group is found to be $M/L_{V}$ = 44 $\pm$ 12 in 
solar 
units. Such a high M/L value is an order of magnitude larger than 
the mass-to-light ratio in the solar neighborhood of the Galaxy. 
This supports the conclusion by Kahn \& Woltjer (1959) that the mass 
of the Local Group is dominated by invisible matter.

\section{NEIGHBORHOOD OF THE LOCAL GROUP}
   The Local Group is situated in the outer reaches of the Virgo
supercluster. The nearest neighbor of the Local Group is the small
Antlia group (van den Bergh 1999a). This tiny cluster is 
located 
at a distance of only 1.7 Mpc from the barycenter of the Local 
Group, i.e. well beyond the zero-velocity surface of the Group. The 
Antlia group has a mean radial velocity of +114 $\pm$ 12 km s$^{-1}$. 
The 
number of galaxies brighter than M$_{V}$ = -11.0 in the Local Group 
is 
22, compared to only four such objects in the Antlia group. 
However, because the Antlia group contains no supergiant galaxies 
like M31 and the Milky Way system, its integrated luminosity is 
$\sim$150 times smaller than that of the Local Group. Since the 
Local
Group is a relatively small cluster, this result suggests that
clusters of galaxies have a range in luminosities (and masses?) 
that extends over at least four orders of magnitude.

   The Centaurus group (van den Bergh 2000b), at a 
distance of $\sim$3.9 
Mpc, which contains M 83 (NGC 5236) and Centaurus A (NGC 5128), is 
the nearest massive cluster. If one assumes NGC 5128 and NGC 5236 to 
be members of a single cluster one obtains a total virial mass of 
1.4 x 10$^{13}$ M$\odot$ and a zero-velocity radius of 2.3 Mpc for the 
Centaurus group. In other words the zero-velocity surfaces of the 
Centaurus group and the Local Group would almost touch each other. 
However, Karachentsev et al. (2002) conclude that 
the galaxies 
surrounding NGC5128 and M83, respectively, actually form dynamically 
distinct clusterings. If that is indeed the case then the total mass 
of these clusters is reduced to only 3 x 10$^{12}$ M$\odot$, and the 
radius of the zero velocity surface around Cen A is less than 1.3 Mpc. 

   A second massive nearby cluster contains IC342 and the highly
obscured elliptical Maffei 1. McCall (1989) concluded 
that ``it is 
likely that IC342 and Maffei 1 had a significant impact on the past 
dynamical evolution of the major members of the Local Group''. More 
recently Fingerhut et al. (2003ab) have, however, 
found that the 
Galactic absorption in front of Maffei 1 is lower than was 
previously
believed. As a result the distances to Maffei 1, and its companions, 
are too large for these objects to have had significant dynamical 
inetractions with individual Local Group members since the Big Bang.

   If the peculiar velocities of galaxies are induced by  
gravitational interactions, then one might have expect massive 
field galaxies to have a lower velocity dispersion than dwarfs. 
Data on the nearby field (Whiting 2003) do not appear 
to support
this expectation. Whiting finds the mean radial velocity
dispersion among field galaxies within 10 Mpc to be 113 km s$^{-1}$.
A much lower dispersion of $\sim$30 km s$^{-1}$ for the local Hubble 
flow 
has, however, been found by Karachentsev et al. 
(2002). It is noted 
in passing that the former value appears to be significantly 
larger than the radial velocity dispersion of 61 $\pm$ 8 km 
$s^{-1}$ that 
Courteau \& van den Bergh (1999) found within 
the Local Group itself.   

\section {SUBCLUSTERING IN THE LOCAL GROUP}

  To first approximation the Local Group is a binary system with
massive clumps of galaxies centered on M31 and on the Galaxy. van 
den Bergh (2000, p. 290) estimates a mass M(A) = (1.15-1.5) x 
10$^{12}$ M$\odot$  for the Andromeda subgroup of the Local Group, 
compared
to M(G) = (0.46-1.25) x 10$^{12}$ M$\odot$  for the Galactic 
subgroup. More
recently Sakamoto, Chiba \& Beers (2003) have given somewhat
higher mass estimates. If Leo I is included they find M(G) =
(1.5-3.0) x $10^{12}$ M$\odot$, compared to M(G) = (1.1-2.2) x 
10$^{12}$ M$\odot$  if
it is assumed that Leo I is not a member of the Galactic subgroup.  
Recent proper motion observations by Piatek et al. 
(2002) suggest
that the Fornax dwarf spheroidal galaxy may not, as had previously
been thought, be a distant satellite of the Galaxy. Instead the
data appear to indicate that Fornax is a free-floating member
of the Local Group that is presently near periGalacticon. Within
each of the two main Local Group subclusters there are additional 
subclumps such as the M31 + M32 + NGC205 triplet, the NGC147 + 
NGC185 binary, and the LMC + SMC binary. For the entire Local Group 
one finds, at the 99\% confidence level, that low-luminosity early-type dSph galaxies are more concentrated in subclumps than are 
late-type dIr galaxies. In other words most dIr galaxies appear to 
be free-floating members of the Local Group, whereas the majority 
(but not all) dSph galaxies seem to be directly associated with 
either M31 or the Galaxy. It is not yet clear if the mean ages of 
the stellar populations in dSph galaxies are themselves a function 
of location. van den Bergh (1995) has tentatively 
suggested that 
there is some evidence to suggest that star formation in 
dSph galaxies in the dense regions close to M31 and the Galaxy 
might typically have started earlier than star formation in remote
dSph galaxies.
The data that are shown in Table 2 show a strong correlation 
between the morphological types of faint galaxies with $M_{V} > -16.5$ 
(i.e. objects fainter than M32) and their environment. This 
dependence was first noticed by Einasto et al. (1974). 
Almost all 
Sph + dSph galaxies are seen to be associated with the two dense 
subclusters within the Local Group, whereas most Ir galaxies appear 
to be more-or-less isolated Group members.
It seems quite possible possible (cf. Skillman et al. 
2003) that the faint dIr and dSph galaxies have similar 
progenitors, and that the differences between them that are observed 
now are due to environmental factors that favored gas loss from those 
dwarfs that occurred in dense environments, i.e. near giant 
galaxies.

Within the Local Group the M31, M32, M33 subgroup has a total
luminosity L$_{V}$ = 3.0 x 10$^{10}$ L$\odot$, which is 
significantly greater than that of the subgroup centered on the Milky 
Way System which has 
L $_{V}$ = 1.1 x $10^{10}$ L$\odot$. These luminosities of the M31 
and Galactic 
subgroups account for 71\% and 24\% of the total Local Group 
luminosity, 
respectively. It should, however, be noted that some uncertainty in 
the luminosity ratio of M31, to that of the Galaxy, is introduced by 
the fact that both of these systems are viewed edge-on. As a result the
internal absorption corrections (which may be quite large) are 
uncertain. Nevertheless, the notion that M31 is more massive than the 
Galaxy receives some support from the observation that M31 appears to 
have 450 $\pm$ 100 globular clusters, compared to only 180 $\pm$  20 
such 
clusters associated with the Galaxy (Barmby et al. 
2000). Furthermore 
(Freeman 1999) the bulge mass of M31 is 3.6 x 
10$^{10}$ M$\odot$, which is almost twice as large as the 2 x 10$^{10}$ 
M$\odot$  mass of the Galactic bulge. 
On the basis of these results one might have expected the total mass 
of the M31 subgroup of the Local Group to be two or three times larger 
than that of the Galaxy subgroup. Surprisingly this does not appear 
to be the case Evans \& Wilkinson (2000). 
Using radial velocity 
observations Evans et al. (2000) conclude that ``There 
is no dynamical
evidence for the widely held belief that M31 is more massive - it may 
even be less massive''. More recently Gottesman et al. 
(2002) have also
concluded from dynamical arguments that the mass of M31 ``is unlikely 
to be as great as that of our own Milky Way''. These authors even make 
the heretical suggestion that M31 might not have a massive halo at all! 
Either the well-known perfidity of small-number statistics has mislead 
us about the relative masses of M31 and the Galaxy, or the mass-to-light ratio in the Milky Way system is much higher than that of the
Andromeda galaxy. If the latter conclusion is correct then one would 
have to accept the existence of significant galaxy-to-galaxy variations 
in the ratio of visible to dark matter among giant spirals. 

\section{THE HALOS OF M31 AND THE GALAXY}

    It has been known for many years (van den Bergh 1969) that the
halo of M31 contains some globular clusters that are much more 
metal rich than those that occur in the Galactic halo. Perhaps the 
best know example of such an object is the luminous globular Mayall 
II.
Furthermore the color-magnitude diagrams for individual M31 halo 
stars
(Mould \& Kristian 1986, Pritchet \& van den Bergh 1988, Durrell, 
Harris \& Pritchet 1994) all show that (1) halo stars have a 
wide range in
metallicity, and (2) the mean metallicity of stars in the halo of
M31 is surprisingly high. The mean values of [Fe/H] for M31 halo 
stars
that were obtained by Mould \& Kristian, by Pritchet \& van den Bergh 
and 
by Durrell, Harris \& Pritchet are $<[Fe/H]>$ = -0.6, -1.0 and 
-0.6,
respectively. [It is noted in passing that the halo of M31 does contain 
a metal-poor component which includes clusters such as Mayall IV,
some RR Lyrae variables, and non-variable horizontal-brach stars 
(Sarajedini \& Van Duyne 2001). The 
observation that the stars in the 
halo of M31 appear, on average, to be so much more metal-rich than than 
those in the Galactic halo (Durrell, Harris \& Pritchet 
1994) suggests 
that these two giant spiral galaxies had quite different evolutionary 
histories. The higher metallicity of M31 halo stars indicates that the 
building blocks of the Andromeda halo had much higher masses than 
those of the Galactic halo. Simulations of Murali et al. (2002)
show that a significant fraction of the mass that was originally in 
such merging ancestral galaxies will end up as intergalactic debris,
and presumably also in an extended halo of the final merged object. 
   An independent check on the metallicities of M31 giants is 
provided by the recent spectroscopic observations undertaken by 
Reitzel \& Guhathakurta (2002). These authors find 
that their spectra 
of stars in the halo of M31 have a mean metallicity $<[Fe/H]>~$= 
-1.3. 
This value is significantly lower that that derived from photometric
observations of stars in the halo of the Andromeda galaxy. A possible 
explanation for this difference is that insufficient correction 
was made for the fact that old metal-rich red giants are fainter 
(and
therefore more difficult to observe) than are old metal-poor giants.
Non homogeneity of the Andromeda halo might also have contributed
to the observed difference in the mean metallicities derived from 
photometry and from spectroscopy. For example, Reitzel \& Guhathakurta 
find four stars of solar metallicity in the halo of M31. They suggest 
that these objects might represent metal-rich debris from an accretion 
event. Other evidence for such accretion events is provided by 
Ibata et al. (2001) and Ferguson et al. 
(2002). Recently Yanny et al. (2003)
have also found possible evidence for such a tidal stream in the
Galaxy. This stream is located beyond the plane of the Milky Way at a 
distance of $\sim$20 kpc from the Galactic center.  

\begin{figure}
\vspace{16.5pc}
\caption{Profile of M31 derived from star counts.  The figure shows 
that an R$^{1/4}$ law provides a reasonable fit to the observations.  
This favors the interpretation that M31 was formed from violent 
relaxaton following mergers.}
\label{fig1}
\end{figure}
                 
An interesting clue regarding the origin of the difference 
between the Andromeda and the Galactic halos is provided by the 
observations of Pritchet \& van den Bergh (1994)] [see Figure 1] which show that M31 has an R$^{1/4}$ profile out to $\sim$ 20 kpc from its 
nucleus. Such a structure is likely to have resulted from violent relaxation. 
This suggests that the overall morphology of the Andromeda galaxy 
was determined by violent relaxation resulting from the early merger 
of two (or more) massive metal-rich ancestral objects. On the 
other hand the main body of the Milky Way system may have been
assembled \`{a} la Eggen, Lynden-Bell \& Sandage (1964)], with its metal-poor halo forming \`{a} la Searle \& Zinn (1978) 
via late infall and 
capture of ``small bits and pieces''. However, it appears that these 
fragments differed in a rather fundamental way from those that 
produced dwarf spheroidal galaxies. Shetrone, C\^{o}t\'{e}, \& 
Sargent (2001)
find that dSph galaxies are iron-rich and have  0.02 $\leq$ [$\alpha$ 
/Fe] $\leq$ 0.13,
compared to typical Galactic values of [$\alpha$/Fe] $\sim$ 0.28 dex 
over the same 
range of metallicities. This shows that the bulk of the Galactic halo 
stars cannot have formed in dwarf spheroidal galaxies that 
subsequently
disintegrated. In particular Fulbright (2002) finds 
that less than 10\% 
of local metal-poor stars with $[Fe/H] < -1.2$ have alpha-to-iron
abundance ratios similar to those found in dSph galaxies. More 
generally
Tolstoy et al. (2003) conclude that the observed 
element abundance 
patterns make it difficult to form a significant proportion of the 
stars observed in our Galaxy in small galaxies that subsequently merged 
to form the disk, bulge, and inner halo of the Milky Way. 
   Bekki, Harris \& Harris (2003) have 
studied the distribution of
stars of various metallicities after the merger of two spirals. 
However, their model is not likely to be applicable to the early 
merger of the ancestral objects of M31. The reason for this is that 
extended disks would be destroyed by frequent tidal interactions at 
large look-back times. {\it Hubble Space Telescope} images show that 
galaxies with obvious disks mostly have $z < 1.5$, whereas the vast
majority of the objects seen at $z > 1.5$ appear to have either 
compact or chaotic morphologies (van den Bergh 2002b). 
So the Bekki et 
al. model, in which extended disks merge, is probably inappropriate 
for galaxies at $z > 1.5$, i.e. for mergers that took place more than 9 
Gyr ago.
   In their pioneering study of the metallicities of individual 
stars
in galactic halos Mould \& Kristian (1986)
also observed stars in the
halo of the late-type spiral M33. Their color-magnitude diagram
suggested that the halo of M33 was very metal-poor and had $<[Fe/H]>~= -2.2$. This value is more than an order of magnitude lower than that for stars in the halo of the Andromeda galaxy.

The discussion given above may be summarised by saying that M31
may have formed from the early merger of the two or three most
massive galaxies in the core of the Andromeda subgroup of the
Local Group. The less-massive Andromeda companions, such as M32
and NGC205, may represent objects in the core of the Andromeda
subgroup which had such low masses that they managed to survive individually.

\section{HISTORY OF GLOBULAR CLUSTER SYSTEMS}

Due to differences in evolutionary history the halo of M31 
contains relatively metal-rich globular clusters, whereas the 
Galactic halo does not. Another difference between the M31 and
Galactic globular cluster systems has been noted by Rich et al. 
(2002) who find that M31 does not appear to contain globular
clusters with extremely blue horizontal branches such as M92 in 
the Galactic halo. An additional example of major differences 
between globular cluster systems is provided by M33 and the LMC. 
These two late-type disk galaxies have comparable luminosities 
(M33 ~~M$_{V}$ = -18.9 , LMC~~ M$_{V}$ = -18.5) but radically 
different globular 
cluster systems. Surprisingly the LMC globulars, which are both very 
old and quite metal-poor, appear to have disk kinematics (Schommer 
et al. 1993). On the other hand the metal-poor globular 
clusters 
associated with M33 (Schommer et al. 1991) seem to 
have halo-like 
kinematics. Sarajedini et al. (1998 found that 
eight out of 10 
globular clusters in their M33 halo sample had significantly redder 
horizontal branches than Galactic globulars of similar metallicity. 
This difference might be interpreted as a second parameter effect.
Alternatively, and perhaps more plausibly, the M33 globulars may
be a few Gyr younger than their Galactic counterparts. On the latter
interpretation the M33 halo globular clusters exhibit an unexpectedly 
large age dispersion of $\sim$3-5 Gyr. It is presently a mystery why 
the 
M33 halo globular clusters would have formed a few Gyr later than 
typical Galactic halo and LMC disk globulars. Some light might 
eventually be shed on these questions by observations of the radial 
velocities of RR Lyrae stars in the LMC, and perhaps also in the 
near future, of RR Lyrae stars in M33. 
The main differences between the globular cluster systems of M33 
(Sc) and the Galaxy (Sbc) could perhaps be understood (van den 
Bergh 2002b)
 by assuming that late-type galaxies take significantly longer 
to arrive at their final morphology than do spirals of earlier 
morphological types. However, the great age of the LMC cluster system 
appears to conflict with this simple explanation. The observations of 
Mould \& Kristian (1986)] appear to show that 
the field stars in the 
halo of M33 are extremely metal-poor and have $<[Fe/H]>~= -2.2$. 
It
would be important to confirm this result by new photometry and to
compare this value with the mean metallicity of stars in the outer
halo of the Large Magelanic Cloud.

   Not unexpectedly the majority of Local Group dwarf galaxies are 
surrounded by small families of metal-poor globular clusters. 
However, it is puzzling that the Sagittarius dwarf has one globular 
cluster companion (Terzan 7, [Fe/H] = -0.36) that is quite 
metal-rich.
How could such a relatively high metallicity have been built up
within a dwarf galaxy?  The only other Galactic halo (R$_{gc} >$ 10 
kpc) 
globular clusters that are known to have  metallicities higher than 
[Fe/H] = -1.0 are Pal 1 and Pal 12. The latter object is itself 
suspected of also being associated with the disintegrating Sagittarius  
dwarf (Irwin 1999). This speculation is supported by 
the observations 
of Mart\'{i}nez-Delgado et al. (2002) who have 
found that Pal 12 is
possibly imbedded in tidal debris of the Sagittarius dwarf. 
   Rosenberg et al. (1998ab) have also shown that 
the cluster Pal 1 is
significantly younger than most other Galactic globular clusters. It 
is presently not clear which kind of evolutionary scenarios would allow 
halo clusters like Pal 1 and Pal 12 to attain such relatively high
metallicities. From its present luminosity and morphological type the 
dwarf elliptical galaxy M32 would have been expected to be embedded in 
a swarm of 10 - 20 globular clusters. It is therefore puzzling that 
not a single globular cluster appears to be associated with this 
galaxy.
Perhaps some of the innermost M32 clusters were dragged into its
compact luminous nucleus by dynamical friction. Also loosely bound  
outer globulars, that were originally associated with M32, might have 
been detached by tidal interactions with the main body of M31. Such 
detached M32 clusters would remain in the halo of M31 and might be 
recognized by being unusually compact. It would be very worthwhile 
to undertake a systematic search for such M32 clusters with small 
R$_{h}$ 
values in the halo of M31.

For the vast majority of galaxies the specific globular cluster 
frequency S (Harris \& van den Bergh 1981) is 
less than 10.  However, the Fornax dwarf, which has 5 globulars 
associated with it, has $13 < S < 26$. Recent work by Kleyna et 
al. (2003) appears to indicate that the Ursa Minor dwarf may 
have an even higher S value.  If a dynamically cold clustering of stars 
that these authors find in UMi is a globular cluster (or a 
disintegrated cluster) then S $\sim$400 for the UMi system.  Taken at 
face value this result suggests, that the fraction of the light of dwarf 
spheroidals, that is in the form of globular clusters, may be much higher 
in dwarf spheroidal galaxies than it is in more luminour (massive) 
systems.
 
If the Milky Way system had collapsed in the fashion advocated
by Eggen, Lynden-Bell \& Sandage (1962), then one would have 
expected the stars and globular clusters in the Galactic halo to
exhibit a radial metallicity gradient. On the other hand the halo 
of the Milky Way system would not be expected to have such a 
metallicity gradient if, as envisioned by Searle \& Zinn 
(1978), it 
had formed by the accretion of many ``bits and pieces''. Using the 1999 
version of the globular cluster catalog of Harris (1996), van den 
Bergh (2003), found a possible hint for the existence of 
such a radial
metallicity gradient among Galactic halo (R$_{gc} > $10 kpc) globular 
clusters. However, the reality of this a gradient is not supported by 
the more recent data contained in the 2003 version of Harris's 
catalog.
On the other hand the clusters in the main body of the Galaxy, i.e. 
those with R$_{gc} < $10 kpc, do appear to show a radial abundance 
gradient. Globulars at R$_{gc} < 4.0$ kpc are, on average, found to
be more metal-rich than those having 4.0 $\leq$  R$_{gc} < 10$ kpc. A 
Kolmogorov-Smirnov test shows that there is only a 4\% probability 
that the metallicities of these inner and outer globular clusters 
samples were drawn from the same parent population. Taken at face 
value the existence of a Galactic metallicity gradient between 4 kpc 
and 10 kpc favors the suggestion that the ELS model provides an 
adequate description of the formation of the main body of the Galactic 
halo, whereas the SZ model predictions agree with the observed lack of 
a metallicity gradient in the region with R$_{gc} > $10 kpc.

   It is a curious (and unexplained) fact that the distribution
of flattening values of globular clusters differs significantly from 
galaxy to galaxy. Both open and globular clusters in the LMC are, for 
example, on average more flattened than their Galactic counterparts.
Furthermore (van den Bergh 1983) the luminous clusters in 
the Large 
Cloud are typically more flattened than are the less luminous ones. 
Finally it is of interest to note that the most luminous globular in 
many Local Group galaxies also seem to be among the most flattened. 
Examples are the globular Mayall II ($\epsilon$  = 0.22) in M31, 
$\omega$   Centauri 
($\epsilon$ = 0.19) in the Galaxy, and NGC1835 ($\epsilon$ = 0.21) 
in the LMC. 

\section{THE LUMINOSITY FUNCTION OF THE LOCAL GROUP}

   The presently known members of the Local Group exhibit a
flat luminosity function with slope $\alpha$ = -1.1 $\pm$ 0.1 
(Pritchet \& van den Bergh 1999). This value is 
significantly lower than the
slope $\alpha$ = -1.8 that is  predicted by the cold cold dark 
matter
theory (Klypin et al. 1999). The low frequency of faint 
Local
Group dwarfs has been confirmed in a recent hunt for new faint
Local Group members by Whiting, Hau \& Irwin (2002). This observed
deficiency of faint LG members suggests that many of the
progenitors of low mass galaxies were destroyed before they had
a chance to form significant numbers of stars. Alternatively it
might be assumed that the missing faint galaxies can be 
identified with compact high-velocity clouds. However, Maloney \& 
Putman (2003) have recently shown that such 
objects, if they
were located at distances of $\sim$1 Mpc, would be largely ionized.
These authors therefore conclude that the compact high-velocity
clouds are not at cosmologically significant distances, but that
they are instead associated with the Galactic halo. All attempts to 
search for evidence of star formation in compact high-velocity 
clouds (e.g. Simon \& Blitz 2002) have so far 
remained unsuccessful. It is therefore the deficiency of faint Local Group 
members is real. Figure 2 appears to show (van den Bergh 2000a, p. 
281) that the luminosity function of dSph galaxies in the Local Group is steeper 
than that for dIr galaxies. This conclusion should, however, be 
regarded as provisional because a Kolmogorov-Smirnov test shows that 
the difference between the luminosity distributions of Local Group 
dSph and dIr galaxies is only significant at the 75\% confidence 
level. If the luminosity function of dSph galaxies is, indeed, 
steeper than that for dIr galaxies, then future discoveries are 
most likely to turn up very faint dSph (rather than dIr) members 
of the Local Group. It would clearly be very important to undertake
sky surveys in two (or more) colors to search for the signatures of
the color-magnitude diagrams of extremely faint (and so far 
undiscovered) resolved dwarf members of the Local Group.

\begin{figure}
\vspace{16.5pc}
\caption{Luminosity function of the Local Group.  The data suggest, but 
do not prove, that the dIr luminosity distribution is less steep than that 
for dSph galaxies.}
\label{fig2}
\end{figure}

Three distinct explanations might be invoked to account for the 
apparent excess of faint dSph galaxies among presently known 
Local Group members: (1) Perhaps gas was more likely to escape from 
faint (low-mass) galaxies than from more massive objects. As a 
result low-mass galaxies would most often end up as gas free 
dSph galaxies. Alternatively (2) the mass spectrum with which 
galaxies form might depend on environmental density in such 
a way that high density regions (i.e. the neighborhood of M31 and 
the Galaxy) form a larger fraction of low mass objects (the 
majority of which end up as dSph galaxies). The latter assumption 
would be consistent with the work of Trentham et al. 
(2001) and 
Trentham \& Tully (2003) who found that the 
galaxian luminosity 
function of the dense Virgo cluster is much steeper than that for 
less dense clusters such as the Ursa Major cluster and the Local 
Group. On the other hand the view that dense regions produce 
galaxies with steep galaxian luminosity functions appears to 
conflict with the result of Sabatini et al. (2003), 
who find that 
the Virgo cluster luminosity function seems to be steeper in the 
low density outer regions of the Virgo cluster than it is in the 
high density core of this cluster. This observation might, however, 
be accounted for by assuming that tides preferentially destroy
fragile dwarfs in the cores of dense clusters. Finally, (3) and 
perhaps most plausibly, the gas in the progenitors of the missing 
dwarfs might have been photoevaporated during reionization.

\section {MORPHOLOGICAL EVOLUTION OF LOCAL GROUP MEMBERS}

   It is difficult to tease out information on the morphological
evolution of galaxies from the distribution of stars of various
ages. The central bulges of giant spirals, such as M31 and the 
Galaxy, are dominated by old stars. This supports the notion 
that these objects were built up inside out, with their bulges
forming first and the disk possibly being accreted at a later
time. It would be very important to establish how old the first
(presumably quite metal-poor) generation of Galactic disk stars 
is. This problem is made more intractable because such thin disk
stars have to be disentangled from stars that are physically 
associated with, and embedded within, an older thick disk 
population. In fact, tidal interactions might pump energy into
(and hence thicken) an initially thin disk of very metal-poor
stars. {\it Hubble Space Telescope} observations of galaxies at at 
large look-back times suggest (van den Bergh 2002b) that 
most
disk star formation occurs at $z < \sim$1.5, i.e. during the last 
9 Gyr. One reason for the paucity of disks at larger redshifts is, 
presumably, that such extended structures would be destroyed by 
tidal forces during the frequent encounters between galaxies at 
high redshifts. From two slightly metal-poor stars with disk 
kinematics Liu \& Chaboyer (2000) find a thin 
disk age of 9.7 $\pm$ 0.6 
Gyr. Such an age is consistent with the ages of spiral disks that 
are inferred from the fact that the HST images of distant galaxies 
start to show obvious disks at z $\sim$1.5. 

   Since bars are generally assumed to have formed from global 
instabilities in disks one would not expect to see barred galaxies
at $z > 1.5$. If bars can not form from from initially chaotic
protodisks then bar formation might be delayed to even smaller
look-back times. This suspicion appears to be confirmed by
observations (van den Bergh et al. 1996, 2002) which 
seem to
suggest that the frequency of barred galaxies declines precipitously 
beyond redhifts of  z $\sim$0.7. If this conclusion is correct then 
one would expect the Bar of the LMC to be younger than 6 Gyr. This 
conclusion is consistent with (but not proved by) the observation
(Smecker-Hane et al. 2002) that bursts of star 
formation  occurred 
in the Bar of the Large Cloud 4 - 6 Gyr and 1 - 2 Gyr ago.

\section {THE HISTORY OF STAR AND CLUSTER FORMATION}
   Almost all Local Group galaxies are found to contain some
very old stars like RR Lyrae variables. This shows that these
galaxies started to form stars quite early in the history of the 
Universe, i.e. more than $\sim$10 Gyr ago. The best candidate for a 
``young'' Local Group member is the dwarf irregular Leo A. However, 
Dolphin et al. (2002) have discovered a few RR Lyrae 
variables, which have ages $>9$ Gyr, in this galaxy. Furthermore, 
recent observations by Schulte-Ladbeck et al. 
(2002) show that 
this object also contains some metal-poor red horizontal branch 
stars. This clearly demonstrates that Leo A is not a young galaxy. 
   M32 and many of the Local Group dSph galaxies have not
experienced recent star formation. On the other hand the
gas rich spiral and irregular Group members are still
forming stars at a significant rate. Qualitative data on the
past rate of star formation in such galaxies can be obtained
from both their integrated colors and from the intensity ratios
of various spectral lines. However, the hope that the age
distribution of star clusters in galaxies could provide more detailed 
information on the past rate of star formation has been 
shattered by the work of Larsen \& Richtler (1999, 2000) which 
appears to show that the rate of cluster formation varies as a 
rather high power of the rate of star formation. In other words 
there is not a one-to-one correspondence between the rate of 
cluster formation and the general rate of star formation. In 
the Local Group this phenomenon is beautifully illustrated by 
the difference between the quiescent dwarf irregular IC1613,
which contains few star clusters of any kind (i.e. Baade 1963, p. 
231; van den Bergh 1979), and the Large Magellanic Cloud that 
is presently forming both stars and clusters quite actively. It 
seems likely that the present specific frequency of globular 
clusters in galaxies was mainly determined by their peak rates 
of star formation, with elevated peak rates resulting in high 
present specific cluster frequencies. Only fragmentary information 
is so far available on the luminosity evolution of individual 
Local Group members. Few star clusters in the LMC have ages 
between the 3.2 Gyr age of the oldest open clusters and the $\sim$13 
Gyr 
(Rich, Shara \& Zurek 2001) age of the LMC 
globular clusters. This 
probably means that the Large Cloud experienced a quiescent period 
that extended for $\sim$ 10 Gyr. During this ``dark age'' no violent 
bursts 
of star formation (which could have triggered the formation of star 
clusters) occurred. However, it is quite likely that a trickle of 
star formation (such as that which presently occurs in IC1613) 
continued during the dark ages between 3.2 Gyr and 13 Gyr ago. This
speculation is supported by the data of Da Costa 
(2002) which seem
to show that the metallicity in the LMC increased between the 
beginning and the end of the dark age, i.e. between the termination 
of globular cluster formation $\sim$13 Gyr ago, and the beginning of 
open cluster formation 3.2 Gyr ago. A possibly more complicated 
scenario is hinted at by the work of Smecker-Hane et al. 
(2002) who 
conclude that star formation in the Bar of the LMC was episodic, 
while the rate of star formation remained more or less constant 
within the disk of the Large Cloud. However, an important caveat
is that the rate of star formation in LMC the disk was so low that
the data do not provide strong constraints on the LMC disk star
formation history.

   The Carina dSph galaxy seems to have experienced a major burst 
of star formation 7 Gyr ago (Hurley-Keller, Mateo \& Nemeic 
1998)]. 
However, at maximum this object probably only reached M$_{V}$ $\sim$ 
-16 
making it too faint to have become what Babul \& Ferguson 
(1996) have called a ``boojum''.

   There has been a long controversy among astronomers regarding
the nature (or even the existence of) a fundamental difference 
between open clusters and globular clusters. The present
consensus is that all clusters populations initially formed with 
an power law mass spectrum, and that globular clusters are simply 
the oldest and most massive population component that was best 
able to withstand the erosion caused by the destruction of
lower mass clusters via evaporation, encounters with massive
interstellar clouds and disk/bulge shocks. However, a different
scenario has been proposed by van den Bergh (2001). He 
suggested 
that there have, in fact, been two (perhaps quite distinct) epochs  
of cluster formation. During the first of these globular clusters 
might have formed as halo gas was being compressed by shocks that
were driven inwards by ionization fronts generated during cosmic 
reionization at $5 < \sim z < \sim 15$. Such effects would presumably 
be 
greatest in the halos of small protogalaxies that are relatively
easy to ionize. A second generation of massive clusters might have 
formed by the compression (and subsequent collapse) of giant 
molecular clouds that was triggered by heating of the interstellar 
medium induced by collisions between gas rich protogalaxies. A 
similar view has recently been expressed by Schweizer 
(2003) who 
also argues that the first generation of globular clusters formed 
nearly simultaneously from pristine molecular clouds that were heated 
and
shocked by the strong pressure increase that accompanied 
cosmological reionization. Schweizer argues that this hypothesis 
might also account for the similarity of metal-poor globular 
clusters in all types of galaxies and environments. Both van den 
Bergh (2001) and Schweizer (2003) argue 
that second generation 
globular clusters were mainly formed during subsequent collisions and 
mergers between galaxies. 
 
\section {INTERGALACTIC MATTER}
   From dynamical arguments Kahn \& Woltjer (1959)] first showed
that the Local Group can only be stable if it contains a
significant amount of invisible matter. Using radial velocity
observations of Local Group members Courteau \& van den Bergh 
(1999)
have estimated that the Local Group has a total mass of
(2.3 $\pm$  0.6) x 10$^{12}$ M$\odot$, from which the mass-to-light 
ratio
(in solar units) is found to be M/L $_{V}$ = 44 $\pm$ 14. This 
high
value shows that the total mass of the Local Group exceeds that
of the visible parts of its constituent galaxies by an order of 
magnitude. In their 1959 paper Kahn \& Woltjer suggested that this
``missing mass'' in the Local Group might be in the form of hot
(5 x 10$^{5}$ degrees) low density (1 x 10$^{-4}$ protons cm$^{-3}$) gas, 
which
would be difficult to detect observationally. Hui \& Haiman 
(2003)
have recently shown that the thermal history of such gas has
probably been quite complex. In recent years the notion that 
hot gas is responsible for the missing mass in the Local Group
has been overshadowed by the idea that this missing mass is actually
in the form of cold dark matter (Blumenthal et al. 
1984). However, 
recent \textit{Far Ultraviolet Explorer} satellite observations (Nicastro 
et al. 2003) of the absorption lines of O VI (which have 
radial
velocities of only a few hundred km s $^{-1}$) suggest that hot gas
may provide a non-negligible contribution to the missing mass in
the Local Group. Alternatively Sternburg 2003 the 
hot gas clouds 
observed by Nicastro et al. might just be reprocessed metal-enriched
gas, that was ejected from the neighborhood of the Galactic plane in 
supernova driven fountains. If the latter suggestion is correct, 
then one might expect that the hot clouds would be relatively metal
-rich. On the other hand they would be expected to be metal-poor if 
they are composed of hot primordial gas. 

\section {THE END}
   The final fate of the Local Group has been discussed by Forbes 
et al. (2000) who conclude that dynamic friction will 
eventually 
result in the merger of M31 and the Galaxy. This merged object
will resemble an elliptical galaxy with M$_{V}$ $\sim$ -21 that 
contains 
$\sim$ 700 globular clusters. In a Universe that continues to expand 
for
ever (Bennett et al. 2003, Spergel et al. 2003) this object will,
in the distant future, be the only remaining visible object in
the Universe.

\section {SUMMARY}
\begin{itemize}
\item Both the high metallicity of the M31 halo, and the R$^{1/4}$
  luminosity profile of the Andromeda galaxy, suggest that this 
  object might have formed from the early merger, and subsequent
  violent relaxation, of two (or more) relatively massive metal-rich ancestral objects.
\item The main body of the Galaxy may have formed in the manner
  suggested by Eggen, Lynden-Bell \& Sandage (1962), whereas its 
  halo is more likely to have been assembled by accretion of ``bits 
  and pieces'' in the manner first suggested by Searle \& Zinn 
(1978). 
\item It is profoundly puzzling that the old metel-poor globular 
clusters 
  in the LMC appear to have been formed in an early disk, whereas
  the globulars associated with M33 seem to have originated in a 
  slightly younger halo.
\item It is speculated that the oldest generation of globular clusters 
  in the Universe might have formed as halo gas was compressed 
  and heated in shocks that were driven inwards by ionization 
  fronts generated during cosmic reionization. On the other hand 
  second generation globular clusters formed as a result of the 
  heating of molecular clouds during collisions between gas-rich 
  galaxies. It is emphasized that the history of star formation in 
  galaxies is often very different from the history of cluster 
  formation.
\item It is presently not understood how globular clusters like
  Terzan 7 (which is associated with the Sagittarius dwarf) were
  able to attain a relatively high metallicity. Neither do we know
  why the globular clusters associated with some galaxies are so
  much more flattened than are those in others. 
\item It is suggested that the specific globular frequency of 
galaxies
  was mainly determined by the peak rate of star formation during
  evolution.
\item All Local Group galaxies appear to contain a very old 
population
  component, i.e. all nearby galaxies started to form stars just
  after they were formed. In other words there are no truly young
  galaxies in the Local Group. 
\item The Local Group has a mass of (2.3 $\pm$ 0.6) x 10$^{12}$ 
M$\odot$, a luminosity
  L$_{V}$ = 4.2 x 10$^{10}$ L$\odot$  ,and a zero-velocity radius of 
1.18    $\pm$ 0.16 Mpc. Most of the mass and luminosity of the Local 
Group appears to be concentrated in two subgroups that are centered on 
M31 and the Galaxy,
respectively. There is presently a lively controversy about which 
  of these two subgroups is the most massive. If the Galactic subgroup turns out to be more massive than the M31 group, then the
  ratio of dark to visible matter must  differ significantly from group 
  to group. 
\item It is not yet clear if hot low density gas provides a 
significant
  contribution to the total mass of Local Group galaxies.

\end{itemize}

 It is a pleasure to thank Oleg Gnedin, and Eva Grebel for helpful
comments on an early draft of the present paper. I also thank David 
Duncan for preparing the figures.

\clearpage

\scriptsize
\begin{deluxetable}{lcccl}
\tablenum{1}
\tablecaption{Members of the Local Group}

\tablehead{\colhead{Name} & \colhead{Alias} & \colhead{DDO Type} & 
\colhead{D(kpc)} & \colhead{$M_{V}$}}

\startdata

M31        &     NGC 224  &   Sb I-II     &   760     &  -21.2\\
Milky Way  &     Galaxy   &   Sbc I-II:   &   8       &  -20.9\\
M33        &     NGC 598  &   Sc II-III   &   795     &  -18.9\\
LMC        &     ...      &   Ir III-IV   &   50      &  -18.5\\
SMC        &     ...      &   Ir IV/IV-V  &   59      &  -17.1\\
M32        &     NGC 221  &   E2          &   760     &  -16.5\\
NGC 205    &     ...      &   Sph         &   760     &  -16.4\\
IC 10      &     ...      &   Ir IV:      &   660     &  -16.3\\
NGC 6822   &     ...      &   Ir IV-V     &   500     &  -16.0\\
NGC 185    &     ...      &   Sph         &   660     &  -15.6\\
IC 1613    &     ...      &   Ir V        &   725     &  -15.3\\
NGC 147    &     ...      &   Sph         &   660     &  -15.1\\
WLM        &     DDO 221  &   Ir IV-V     &   925     &  -14.4\\
Sagittarius &    ...      &   dSph(t)     &   24      &  -13.3\\
Fornax      &    ...      &   dSph        &   138     &  -13.1\tablenotemark{~7}\\
Pegasus     &    DDO 216  &   Ir V        &   760     &  -12.3\\
Leo A       &    DDO69    &   Ir V        &   800     &  -12.2\tablenotemark{~1}\\
SagDIG      &    ...      &   Ir V        &   1180    &  -12.0\tablenotemark{~2}\\
Leo I       &    Regulus  &   dSph        &   250     &  -11.9\\
And I       &    ...      &   dSph        &   810     &  -11.8\\
And II      &    ...      &   dSph        &   700     &  -11.8\\
Aquarius    &    DDO 210  &    V          &   1025    &  -11.3\\
Pegasus II  &    And VI   &   dSph        &   815     &  -10.5\tablenotemark{~3}\\
And V       &    ...      &   dSph        &   810     &  -10.2\\
And III     &    ...      &   dSph        &   760     &  -10.2\\
Cetus       &    ...      &   dSph        &   775     &  -10.2\tablenotemark{~4}\\
Leo II      &    ...      &   dSph        &   210     &  -10.1\\
Pisces      &    LGS 3    &   dIr/dSph    &   620     &  ~-9.8\tablenotemark{~5}\\ 
Phoenix     &    ...      &   dIr/dSph    &   395     &  ~-9.8\\
Sculptor    &    ...      &   dSph        &   87      &  ~-9.8\\
Tucana      &    ...      &   dSph        &   895     &  ~-9.6\tablenotemark{~6}\\
Cassiopeia  &    And VII  &   dSph        &   690     &  ~-9.5\\
Sextans     &    ...      &   dSph        &   86      &  ~-9.5\\
Carina      &    ...      &   dSph        &   100     &  ~-9.4\\
Draco       &    ...      &   dSph        &   79      &  ~-8.6\\
Ursa Minor  &    ...      &   dSph        &   63      &  ~-8.5\\

\enddata

Uncertain values are marked by a colon (:)

\tablenotetext{1}{from Dolphin et al. (2002)}
\tablenotetext{2}{from Lee \& Kim (2000)}
\tablenotetext{3}{from Pritzel et al. (2002)}
\tablenotetext{4}{from Whiting et al. (1999)}
\tablenotetext{5}{from Miller et al. (2001)}
\tablenotetext{6}{from Walker (2003)}
\tablenotetext{7}{from Majewski et al. (2003)}

\end{deluxetable}

\clearpage

\begin{deluxetable}{llccrr}
\normalsize
\tablenum{2}
\tablecaption{Environments of dwarf members of the Local Group}

\tablehead{&\colhead{Environment} &\colhead{Sph+dSph}
&\colhead{dSph/dIr} &\colhead{Ir}}

\startdata

&Isolated      &     1     &     2     &     6\\
&?             &     1     &     1     &     1\\
&In subcluster &    15     &     0     &     0\\

\enddata

\end{deluxetable}

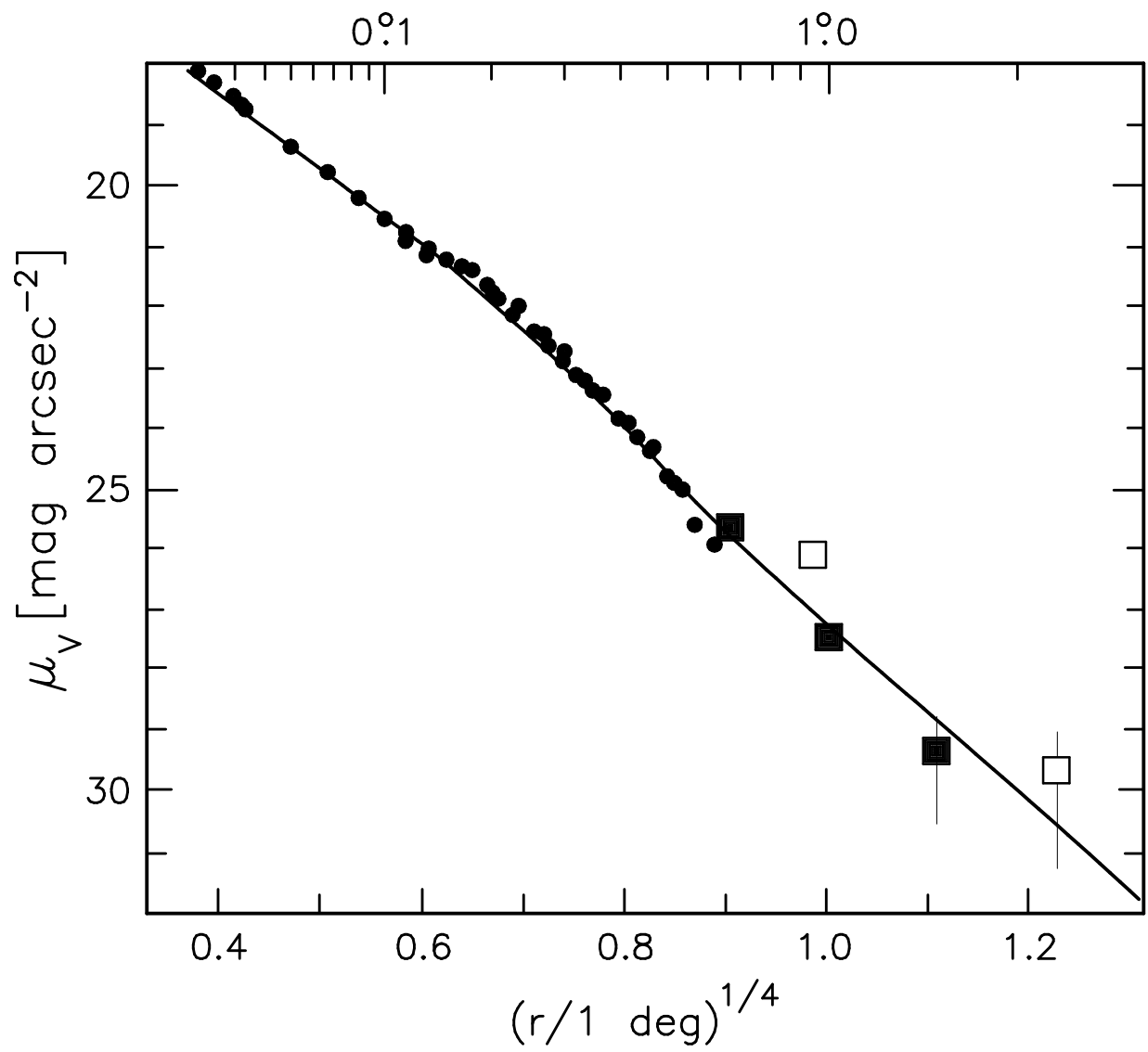
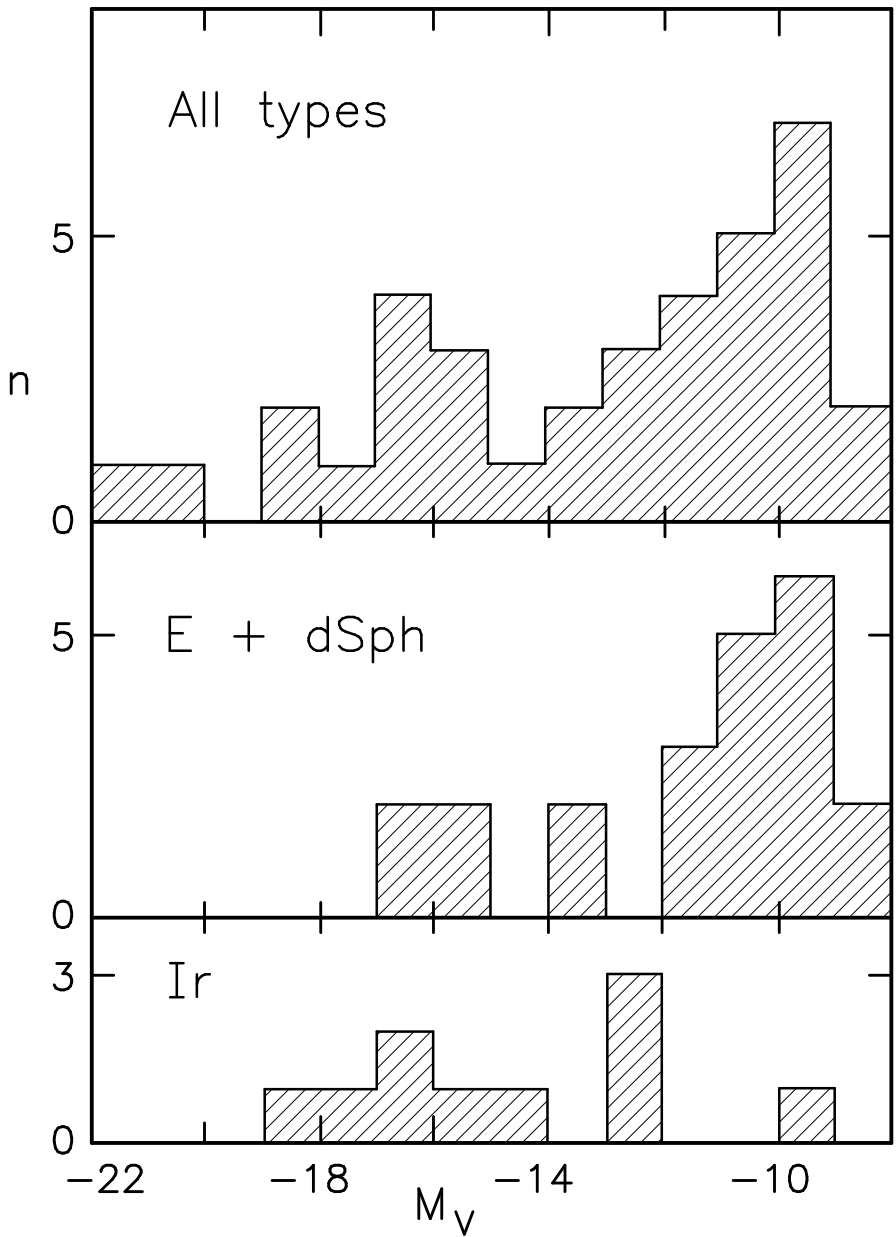
\end{document}